# Allocation and Admission Policies for Service Streams

Michele Mazzucco, Isi Mitrani
*Newcastle University*
*Newcastle upon Tyne, UK*
{*Michele.Mazzucco, Isi.Mitrani*}@ncl.ac.uk

Mike Fisher, Paul McKee
*BT Group*
*Adastral Park, UK*

## Abstract

*A service provisioning system is examined, where a number of servers are used to offer different types of services to paying customers. A customer is charged for the execution of a stream of jobs; the number of jobs in the stream and the rate of their submission is specified. On the other hand, the provider promises a certain quality of service (QoS), measured by the average waiting time of the jobs in the stream. A penalty is paid if the agreed QoS requirement is not met. The objective is to maximize the total average revenue per unit time. Dynamic policies for making server allocation and stream admission decisions are introduced and evaluated. The results of several simulations are described.*

## 1. Introduction

This paper is motivated by the problems arising when attempting to market computer services. A service provider employs a cluster of servers in order to offer a number of different services to a community of users. The users pay for having their jobs run, but demand in turn a certain quality of service. More precisely, a user wishes to submit a specified number of jobs of a given type, at a specified rate; such a collection is referred to as a 'stream'. There is a charge for running a stream (which may depend on the type), and an obligation on the part of the provider that the average waiting time of all jobs in the stream will not exceed a given bound. If that obligation is not met, the provider pays a specified penalty to the user.

Thus, with each service type is associated a *service level agreement* (SLA), formalizing the charge, obligation and penalty corresponding to that type. It is then the provider's responsibility to decide how to allocate the available resources, and when to accept incoming streams, in order to make the system as profitable as possible. Moreover, both the server allocation and the stream admission policies must be dynamic, so that they may react appropriately to changes in demand. This is the problem that we wish to address.

We design and evaluate allocation and admission heuristics that aim to maximize the average revenue received per unit time. They are based on (a) dynamic estimates of traffic parameters, and (b) models of system behaviour. The emphasis of the latter is on generality rather than analytical tractability. Thus, interarrival and service times are allowed to have general distributions with finite coefficients of variation. To handle the resulting models, it is necessary to use approximations. However, those approximations lead to policies that perform well and can be used in real systems.

The revenue maximization problem described here, where both server allocations and stream admissions are controlled simultaneously, does not appear to have been studied before. The most closely related work is by Mazzucco *et al* [8]. In that study, charges, obligations and penalties, and hence admission policies, apply to individual jobs rather than to user streams. The realism of those assumptions can be disputed. Service providers do not normally make QoS promises to every job they accept. Another disadvantage of that model is that a user cannot know in advance which of his/her jobs would be accepted and which would not.

Other authors, such as Villela *et al* [12], Levy *et al* [6], and Liu *et al* [7] have concentrated on optimizing the allocation of server capacity only; admission policies are not considered. Yet the latter have a very significant effect on revenues. Chandra *et al* [3], Kanodia and Knightly [5], Bennani and Menascé [2] and Chen *et al* [4] examine certain aspects of resource allocation and admission control in systems where the QoS criterion is related to waiting or response time. Those studies do not consider the economic issues associated with income and expenditure.

The system model and the associated QoS contracts are described in section 2. The mathematical analysis and the resulting heuristic policies for server allocation and stream admission are presented in section 3. That section also contains a policy optimization algorithm and a different, sim-pler admission policy. A number of experiments where the heuristics are evaluated and compared under different loading conditions, are reported in section 4. Section 5 contains a summary and conclusions.

## 2. The model

The provider has a cluster of $N$ (typically identical) computers, which are used to offer $m$ different services numbered $1, 2, \ldots, m$. A user request for service $i$ is referred to as a 'stream of type $i$'. Such a stream consists of $k_i$ jobs, submitted at the rate $\gamma_i$ jobs per second. If a stream is accepted, all jobs in it will be executed. A stream which has been accepted but not yet completed is said to be 'currently active'. If $L_i$ streams of type $i$ are currently active, then the total current arrival rate of type $i$ jobs is $\lambda_i = L_i \gamma_i$. Denote the squared coefficient of variation of the interarrival intervals by $ca_i^2$.

The service times of type $i$ jobs are i.i.d. random variables with mean and squared coefficient of variation $b_i$ and $cb_i^2$ respectively (the squared coefficient of variation of a random variable is defined as the ratio of its variance to the square of the mean).

Thus, the demand of type $i$ when $L_i$ streams are active is characterized by the 4-tuple

$$(\lambda_i, ca_i^2, b_i, cb_i^2) , \; i = 1, 2, \ldots, m . \quad (1)$$

The *quality of service* experienced by an accepted stream of type $i$ is measured by the observed average waiting time, $W_i$:

$$W_i = \frac{1}{k_i} \sum_{j=1}^{k_i} w_j , \quad (2)$$

where $w_j$ is the waiting time of the $j$th job in the stream (the interval between its arrival and the start of its service). One could also decide to measure the quality of service by the observed average response time, taking also the job lengths into account.

**N. B.** It is worth emphasizing that the right-hand side of (2) is a random variable; its value depends on every job that belongs to the stream. Hence, even if all interarrival and service times are distributed exponentially, one would have to include quite a lot of past history into the state descriptor in order to make the process Markov. This remark explains why some of the approximations that follow are really unavoidable.

Each service-level agreement includes the following three clauses:

1. Charge: For each accepted stream of type $i$, a user shall pay a charge of $c_i$ (this would normally depend on the number of jobs in the stream, $k_i$, and their submission rate, $\gamma_i$).

2. Obligation: The observed average waiting time, $W_i$, of an accepted stream of type $i$ shall not exceed $q_i$.

3. Penalty: For each accepted stream of type $i$ whose $W_i$ exceeds $q_i$, the provider shall pay to the user a penalty of $r_i$.

So, in addition to the 'demand parameters', type $i$ has its 'economic parameters', namely the triple

$$(c_i, q_i, r_i) , \; i = 1, 2, \ldots, m . \quad (3)$$

Within the control of the provider are the 'resource allocation' and 'job admission' policies. The first decides how to partition the total number of servers, $N$, among the $m$ service pools. That is, it assigns $n_i$ servers to jobs of type $i$ ($n_1 + n_2 + \ldots + n_m = N$). Having allocated a server to a pool, it is dedicated to serving jobs of the corresponding type only, until a subsequent re-allocation. The above system model is illustrated in Fig. 1.

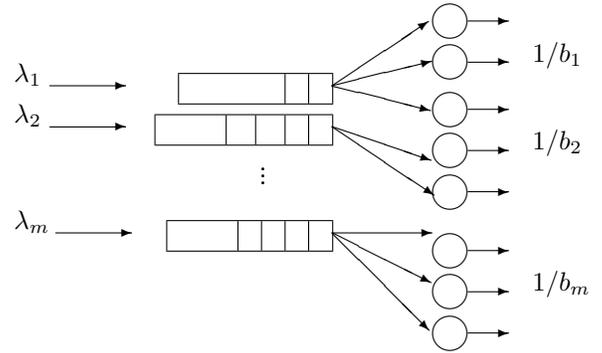

**Figure 1. The model.**
$\lambda_i = L_i \gamma_i$; sq. coeff. of var. $ca_i^2$, $cb_i^2$

The server allocation policy is invoked at stream arrival and stream completion instants. The admission policy is invoked at stream arrival instants. It must decide whether the incoming stream should be accepted or rejected. Of course, the allocation and admission policies are coupled: admission decisions depend on the allocated servers and vice versa. Moreover, both of these policies should respond to dynamic changes in demand. The problem is how to do this in a sensible manner.

We assume that the time it takes to reallocate a server from one queue to another is negligible. That is certainly the case if all services are deployed on all servers, so that a reallocation does not involve a new deployment.

During the intervals between consecutive policy invocations, the numbers of active streams remain constant. Those intervals, which will be referred to as 'observation windows', are used by the controlling software in order to collect traffic statistics and obtain current estimates of the means $1/\lambda_i$ and $b_i$, and coefficients of variation $ca_i^2$ and $cb_i^2$. They are then used by the policies at the next decision epoch.

It is assumed that the observation windows are reasonably large compared to the interarrival and service times, i.e. that enough jobs arrive and are served during a window to provide good traffic estimates and to enable the system to be treated as having reached steady state.

As far as the provider is concerned, the performance of the system is measured by the average revenue, $R$, received per unit time. That quantity can be expressed as

$$R = \sum_{i=1}^{m} a_i[c_i - r_i P(W_i > q_i)] , \quad (4)$$

where $a_i$ is the average number of type $i$ streams that are accepted into the system per unit time and $P(W_i > q_i)$ is the probability that the observed average waiting time of a type $i$ stream, (2), exceeds the obligation $q_i$. The objective of the resource allocation and job admission policies is to maximize the value of $R$.

Note that, although we make no assumptions about the relative magnitudes of the charge and penalty parameters, the more interesting case is where the latter is at least as large as the former: $c_i \leq r_i$. Otherwise one could guarantee a positive revenue by accepting all streams, regardless of loads and obligations.

## 3. Policies

Consider first the dynamic allocation of servers among the different queues. It is proposed to do this roughly in proportion to the observed offered loads, $\rho_i = \lambda_i b_i$, and to a set of coefficients, $\alpha_i$, reflecting the economic importance of the different job types. In other words, when a reallocation decision is to be made, set

$$n_i = \left\lfloor N \frac{\rho_i \alpha_i}{\sum_{j=1}^{m} \rho_j \alpha_j} + 0.5 \right\rfloor \quad (5)$$

(adding 0.5 and truncating is the round-off operation). Then, if the sum of the resulting allocations is either less than $N$ or greater than $N$, adjust the numbers so that they add up to $N$.

This policy will be referred to as the 'Offered Loads' allocation heuristic.

One could use different expressions for the coefficients $\alpha_i$. Experiments have suggested that a good choice is to set $\alpha_i = r_i/c_i$ (the supporting intuition is that the higher the penalty, relative to the charge, the more important the job type becomes). If $r_i = c_i$, then $\alpha_i = 1$ and servers are allocated in proportion to offered loads.

In a practical implementation, a decision to switch a server from one pool to another (taken either at stream arrival or stream completion instant) does not necessarily have to take effect immediately. If a job is being served at the time, it may be allowed to complete before the server is switched (i.e., switching is non-preemptive). Alternatively, one might take the view that preemptive reallocations are preferable.

Now consider the subsystem associated with service $i$, with a given number of streams accepted and hence a given job arrival rate, $\lambda_i$. Suppose that $n_i$ servers have been allocated to queue $i$. The offered load is $\rho_i = \lambda_i b_i$, and the stability condition is $\rho_i < n_i$. If the interarrival intervals and service times could be assumed to be distributed exponentially, then that subsystem could be modelled as an $M/M/n_i$ queue (see, for example, [10]). The average waiting time of a job would be given by

$$w_{M/M/n} = \frac{b_i}{n_i - \rho_i} P(J \geq n_i) , \quad (6)$$

where $J$ is the number of type $i$ jobs present, and so $P(J \geq n_i)$ is the probability that an incoming type $i$ job will have to wait. That probability is given by the Erlang-C formula (or Erlang delay formula)

$$P(J \geq n_i) = \frac{n_i \rho_i^{n_i}}{n!(n_i - \rho_i)} p_0 , \quad (7)$$

with $p_0$ being the probability of an empty type $i$ subsystem:

$$p_0 = \left[ \frac{n_i \rho_i^{n_i}}{n_i!(n_i - \rho_i)} + \sum_{j=0}^{n_i-1} \frac{\rho_i^j}{j!} \right]^{-1} . \quad (8)$$

If the Markovian assumptions are not satisfied, then the appropriate queueing model becomes $GI/G/n_i$, for which there is no exact solution. However, an acceptable approximation for the average waiting time, $\beta_i = w_{GI/G/n}$, is provided by the formula (see Whitt, [13])

$$\beta_i = \frac{ca_i^2 + cb_i^2}{2} w_{M/M/n} , \quad (9)$$

where $ca_i^2$ and $cb_i^2$ are the squared coefficients of variation of the interarrival intervals and service times of type $i$, respectively.

Moreover, when the system is heavily loaded, the waiting time in the $GI/G/n_i$ queue is approximately exponentially distributed (see [13]). Since the variance of the exponential distribution is equal to the mean, the waiting time variance can also be approximated by $\beta_i$. Hence, the observed average waiting time of a stream, which according to (2) involves the sum of $k_i$ waiting times, can be treated as being approximately normally distributed with mean $\beta_i$ and variance $\beta_i/k_i$. That approximation appeals to the central limit theorem and ignores the dependencies between individual waiting times.

Based on the normal approximation, the probability that the observed average waiting time exceeds a given value, $x$, can be estimated as

$$P(W_i > x) = 1 - \Phi\left(\frac{x - \beta_i}{\sqrt{\frac{\beta_i}{k_i}}}\right), \quad (10)$$

where $\beta_i$ is given by (9), (7) and (8), and $\Phi(\cdot)$ is the cumulative distribution function of the standard normal distribution (mean 0 and variance 1). That function can be computed very accurately by means of a rational approximation (see Abramowitz and Stegun [1]).

If $\rho_i \geq n_i$ (violating the stability condition), then it is natural to set $\beta_i = \infty$ and $P(W_i > x) = 1$ for any value of $x$.

The quality of the approximation (10) will depend on how well the implied assumptions are satisfied, namely the load is heavy and there is a large number of jobs per stream (the second of these conditions also ensures that any dependencies between the waiting times within a stream can be neglected). On the other hand, if the system is lightly loaded, then it is not so important to come up with a clever admission policy; all incoming streams would be admitted.

For the following, it will be convenient to indicate explicitly the dependence of (10) on the parameters $\lambda_i$, $k_i$ and $n_i$ by introducing the notation

$$P(W_i > x) = g_i(x; \lambda_i, k_i, n_i), \quad (11)$$

where $g_i(\cdot)$ stands for the right-hand side of (10). The other parameters, $b_i$, $ca_i^2$ and $cb_i^2$ are also involved, but do not need to be acknowledged explicitly.

Consider now an admission decision epoch for service $i$, i.e. an instant when a new stream of type $i$ is offered. The state of subsystem $i$ at that instant is specified by the following values:

(a) The number of streams, $L_i$, currently admitted.

(b) For stream number $t$, the number of jobs already completed, $\ell_t$, and the average waiting time, $u_t$, achieved over those jobs; $t = 1, 2, \ldots, L_i$.

These values, as well as the parameter estimates, are available since traffic is monitored.

Let $v_t$ be the average waiting time over the remaining $k_i - \ell_t$ jobs in stream $t$. A penalty $r_i$ will be payable for that stream if the overall average waiting time exceeds the obligation $q_i$:

$$\frac{u_t \ell_t + v_t(k_i - \ell_t)}{k_i} > q_i, \quad (12)$$

or

$$v_t > \frac{q_i k_i - u_t \ell_t}{k_i - \ell_t}. \quad (13)$$

Denote the right-hand side of (13) by $q_{i,t}$. If the new stream is rejected and no other action is taken, the expected total revenue from the current streams can be estimated as

$$c_i L_i - r_i \sum_{t=1}^{L_i} g_i(q_{i,t}; \lambda_i, k_i - \ell_t, n_i). \quad (14)$$

The alternative is to accept the new stream, possibly in conjunction with a reallocation of servers from other queues to queue $i$. Such a decision would bring in an additional charge of $c_i$, but will also increase the job arrival rate at queue $i$ by $\gamma_i$. There will be a possible penalty to pay for the new stream, and also different probabilities of paying penalties for the existing streams of type $i$, and for the existing streams of the types that lost servers.

Denote by $n'_j$ the new number of servers that queue $j$ would have after a reallocation ($j = 1, 2, \ldots, m$; $n'_1 + n'_2 + \ldots + n'_m = N$). The expected *change* in revenue resulting from a decision to reallocate and accept the new stream can be expressed as:

$$\Delta R = c_i - r_i g_i(q_i; \lambda_i + \gamma_i, k_i, n'_i) - \sum_{j=1}^{m} r_j \sum_{t=1}^{L_j} \Delta g_j(\cdot_t), \quad (15)$$

where $\Delta g_j(\cdot_t)$ is the change in the probability of paying a penalty for stream $t$ at queue $j$. At queue $i$, that change is given by:

$$\Delta g_i(\cdot_t) = g_i(q_{i,t}; \lambda_i + \gamma_i, k_i - \ell_t, n'_i) - g_i(q_{i,t}; \lambda_i, k_i - \ell_t, n_i),$$

while at other queues the change involves only the server reallocation; the arrival rates remain the same:

$$\Delta g_j(\cdot_t) = g_j(q_{j,t}; \lambda_j, k_j - \ell_t, n'_j) - g_j(q_{j,t}; \lambda_j, k_j - \ell_t, n_j).$$

Equation (15) ignores the effect that the admission of a new stream might have on the coefficient of variation of the interarrival intervals. That effect is indeed negligible when the streams are close to Poisson, or when the number of currently active streams is large.

The above discussion suggests that, at stream arrival instants of type $i$ ($i = 1, 2, \ldots, m$), the following policy may be adopted:

1. Invoke the Offered Loads allocation heuristic to determine the numbers of servers, $n'_j$ ($j = 1, 2, \ldots, m$), that the queues would have if the new stream was accepted.

2. Evaluate the expected change in revenue, (15). If it is positive, carry out the server reallocation and accept the incoming stream. Otherwise, reject the new stream and leave the server allocation as it was.

This policy will be referred to as the 'Current State' admission heuristic.

At instances of stream completion, the question of admission does not arise, but that of server reallocation does (since the offered load of one type is reduced). The Offered Load allocation heuristic is invoked and any switching of servers indicated by it is carried out.

The application of Offered Load allocation and Current State admission implies that, if a stream arrives into an otherwise empty system, all servers are allocated to it. Then, if a stream of a different type arrives, several servers are switched to the other queue, etc.

**N.B.** One can easily relax the assumption that all streams of type $i$ have the same job arrival rate, $\gamma_i$, and the same number of jobs, $k_i$. There is no problem in evaluating expressions (15) and making allocation and admission decisions if those quantities vary from stream to stream, as long as each incoming stream announces its arrival rate and number of jobs in advance. The charges, obligations and penalties could also vary from stream to stream.

### 3.1. Policy improvement

The decisions made by the coupled Offered Load allocation and Current State admission heuristics may well be sub-optimal. Consider, for example, an arrival instant of type $i$, when the Offered Load heuristic tries to accommodate the incoming stream. The application of (5) may err either in being too generous to queue $i$ (thus increasing the likelihood of paying penalties in other queues), or not being generous enough (missing out on revenues from queue $i$). Therefore, it could be worthwhile trying to get closer to the 'optimal' server allocation at each decision instant, by carrying out one or more 'policy improvement' steps. At arrival instants, these have the following form.

1. Start with the allocation, $(n'_1, n'_2, \ldots, n'_m)$, that the Offered Loads heuristic would make if the new stream was accepted. Evaluate the corresponding expected change in revenue, $\Delta R$, given by (15).

2. Try the $m - 1$ switches where a server is moved from one of the other queues to queue $i$, and the $m - 1$ switches where a server is moved from queue $i$ to one of the other queues. In each case, evaluate (15) with the new vector $(n'_1, n'_2, \ldots, n'_m)$ and choose the best change; call that value $new\Delta R$.

3. If $new\Delta R > \Delta R$, set $\Delta R = new\Delta R$ and $(n'_1, n'_2, \ldots, n'_m)$ to the corresponding allocation, and repeat step 2; otherwise stop.

4. If $\Delta R$ is positive, carry out the server allocation $(n'_1, n'_2, \ldots, n'_m)$, and accept the incoming stream. Otherwise reject the new stream and leave the allocation as it was.

This procedure implements an optimization algorithm of the 'hill-climbing' variety. It economizes on computation by examining only $2m - 2$ switches, i.e. $2m - 2$ neighbouring allocations at each iteration, rather than all the $m(m - 1)$ possible ones. The intuition is that a switch between a pair of queues neither of which was affected by a change in arrival rate is unlikely to be very advantageous.

A similar policy improvement algorithm can be applied at instants of stream completion of type $i$. Since the arrival rate in queue $i$ has decreased, a sensible server reallocation, e.g. as indicated by the Offered Loads heuristic, would consist of removing a number of servers from queue $i$ and assigning them to other queues. A question then arises whether that number is perhaps too large, or maybe not large enough.

Given the current server allocation, $(n_1, n_2, \ldots, n_m)$, and a proposed reallocation, $(n'_1, n'_2, \ldots, n'_m)$, the expected change in revenue can be estimated by an expression similar to 15:

$$\Delta R = -\sum_{j=1}^{m} r_j \sum_{t=1}^{L_j} \Delta g_j(\cdot_t) , \qquad (16)$$

where

$$\Delta g_j(\cdot_t) = g_j(q_{j,t}; \lambda_j, k_j - \ell_t, n'_j) - g_j(q_{j,t}; \lambda_j, k_j - \ell_t, n_j) .$$

Note that the first two terms in the right-hand side of (15) are now absent, and there is no change in the arrival rates to be considered; the arrival rate in queue $i$ has already been decremented appropriately.

Similar policy improvement steps are carried out at the instants when a stream of type $i$ is completed.

### 3.2. Simpler admission heuristic

The Current State admission heuristic requires a rather detailed knowledge of the states of all active streams in the system. Its computational demands include evaluations of the right-hand side of (10) for every active stream in every queue. It may therefore be desirable to design a simpler heuristic that allows decisions to be taken faster.

Note that the Current State heuristic does not require or make use of the rates, $\delta_i$, at which streams of different types are submitted. Yet one may expect that those parameters could play a role in admission decisions. With that in mind, we propose a heuristic which takes stream submission rates into account.

If streams of type $i$ are submitted at rate $\delta_i$ and each such stream consists of $k_i$ jobs of average length $b_i$ each, then the 'potential' offered load of type $i$ (i.e., if all streams are accepted), is $\phi_i = \delta_i k_i b_i$. Suppose that we allocate the servers to service types in proportion to these loads, using (5) but replacing $\rho_i$ with $\phi_i$. Having fixed those allocations, the different services can be decoupled and considered in isolation of each other.

The proposed admission heuristic is based on a vector of thresholds, $(M_1, M_2, \ldots, M_m)$. If, at the moment when a stream of type $i$ is submitted, there are fewer than $M_i$ active type $i$ streams, the new stream is accepted, otherwise it is rejected. The problem is how to choose those thresholds.

Assuming that the stream submission processes are Poisson, and bearing in mind that the average 'duration' of a type $i$ stream is $k_i/\gamma_i$, we model the number of active streams of type $i$, for a given threshold $M_i$, as the number of calls in an Erlang system with $M_i$ trunks and traffic intensity $\sigma_i = \delta_i k_i/\gamma_i$. The steady state probability, $p_{i,j}$, that there are $j$ active streams of type $i$, is given by (see [10])

$$p_{i,j} = \frac{\sigma_i^j}{j!} p_{i,0} \; ; \; j = 0, 1, \ldots, M_i , \qquad (17)$$

where

$$p_{i,0} = \left[ \sum_{s=0}^{M_i} \frac{\sigma_i^s}{s!} \right]^{-1} . \qquad (18)$$

Since the Erlang model is insensitive to the distribution of call times, we need not worry about the distributions of stream durations.

Now, the average revenue that is obtained per unit time from type $i$ services can be estimated as

$$R_i = \sum_{j=0}^{M_i} p_{i,j} \delta_i [c_i - r_i g_i(q_i; j\gamma_i, k_i, n_i)] , \qquad (19)$$

where $g_i(\cdot)$ is the probability of paying a penalty for a type $i$ stream when there are $j$ such streams active (the job arrival rate is $j\gamma_i$) and $n_i$ servers have been allocated; that probability is given by (10).

The threshold $M_i$ is chosen so as to maximize the right-hand side of (19). When $R_i$ is computed for different threshold values, it becomes clear that it is a unimodal function of $M_i$. That is, it has a single maximum, which may be at $M_i = \infty$ for lightly loaded systems. We do not have a mathematical proof of this proposition, but have verified it in numerous numerical experiments. That observation implies that one can search for the optimal admission threshold by evaluating $R_i$ for consecutive values of $M_i$, stopping either when $R_i$ starts decreasing or, if that does not happen, when the increase becomes smaller than some $\epsilon$. Such searches are typically very fast.

This admission policy will be referred to as the 'Threshold' heuristic. It is a very economical policy to implement, in terms of computational overheads. This is because it is essentially a static policy: the server allocations and the corresponding admission thresholds are computed only once, for a given set of demand parameters. The system is still monitored, and if one or more of those parameters are observed to have changed, the allocations and thresholds are recomputed.

## 4. Results

Several experiments were carried out, using computer simulations. The aim was to evaluate the effects of the server allocation and job admission policies that have been proposed. To reduce the number of variables, the following features were held fixed:

- The cluster consists of 20 servers; two types of services are offered ($n = 20$, $m = 2$).

- The obligations undertaken by the provider are that the average observed waiting time of the jobs in a stream should not exceed their average required service time, i.e. $q_i = b_i$.

- All penalties are equal to the corresponding charges: $r_i = c_i$ (that is, if the average waiting time exceeds the obligation, the user that submitted the stream gets his or her money back).

The first experiment attempts to quantify the extent to which the use of a sensible admission policy (the Current State heuristic) can improve revenues, compared with not having a policy and accepting all submitted streams. The demand parameters of type 1 streams are: $\gamma_1 = 0.2$ (job arrival rate), $b_1 = 10$ (average service time), $k_1 = 50$ (number of jobs in a stream), $c_1 = 100$ (charge and penalty). Type 1 streams are submitted at rate $\delta_1 = 0.02$. The corresponding parameters for type 2 are $\gamma_2 = 0.4$, $b_2 = 5$, $k_2 = 50$ and $c_2 = 200$. However, the offered load of type 2 is increased in different runs, by reducing the average interval between stream submissions from 125 down to about 25 (that is, the submission rate $\delta_2$ increases from 0.008 to 0.04). All interarrival, service and inter-stream intervals are distributed exponentially.

In Fig. 2, the revenues are plotted against $\delta_2$. Each point corresponds to to a simulation run of 110000 time units, divided into 11 portions of 10000 time units each for the purpose of computing a 95% confidence interval. As one might expect, at light loads there is not much difference between having an admission policy and not having one, since nearly all streams are accepted anyway. However, when the system becomes more heavily loaded, the lack of an admission policy begins to have an increasingly significant effect. Whereas the revenues obtained by the Current State heuristic continue to increase throughout with a roughly constant slope, those of the 'Admit all' policy increase more slowly at first, and then quickly drop to near 0.

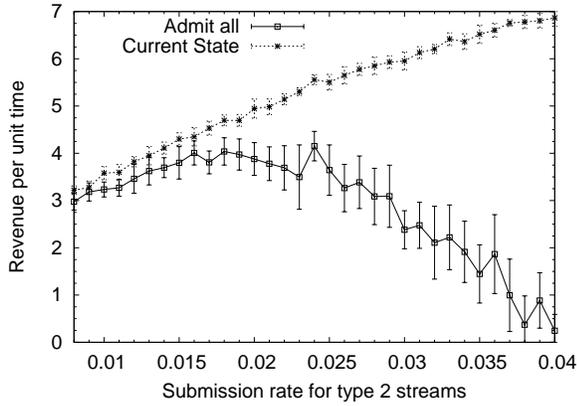

**Figure 2. Benefits of operating an admission policy.**

The next experiment aims to compare the performance of the Current State and Threshold admission heuristics. The setting is the same as before, with the same demand parameters and lengths of simulation runs.

The results are illustrated in Fig. 3. The simple Threshold heuristic performs almost as well as the more complicated Current State one. The revenues obtained are quite close and the confidence intervals are of similar size. This suggests that the Threshold heuristic might be a suitable choice for practical applications.

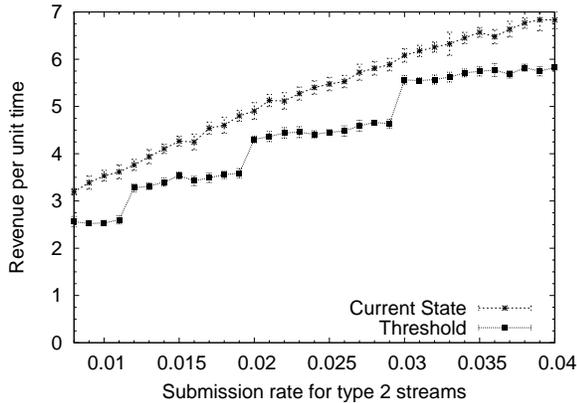

**Figure 3. Comparison of different heuristics.**

The third experiment tries to evaluate the effect of service time variability on performance. As well as the exponentially distributed service times ($cb_i^2 = 1$), the model was run with constant service times ($cb_i^2 = 0$) and with hyperexponential service times ($cb_i^2 > 1$). The average service times we kept the same as before, $b_1 = 10$ and $b_2 = 5$. The hyperexponential distribution had two phases: type 1 service times had a mean of 2 with probability 0.8 and a mean of 42 with probability 0.2; for type 2, the means were 1 with probability 0.8 and 21 with probability 0.2. The corresponding squared coefficients of variation are $cb_1^2 = cb_2^2 = 6.12$.

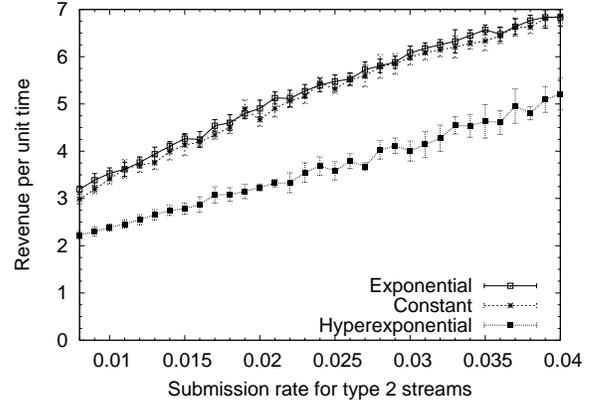

**Figure 4. Service times with different coefficients of variation.**

One expects the performance to deteriorate when the variability of the demand increases, since the system becomes less predictable and it is more difficult to make correct admission decisions. In fact, Fig. 4 shows that almost identical revenues are obtained when service times are distributed exponentially, as when they are constant. However, in the hyperexponential case, which was deliberately chosen to have very high variability, the revenues are indeed lower.

Next, we experimented with a system where there are four different service types. The job arrival rate and the average service time for types 1, 2 and 3 are $\gamma = 2$ and $b = 1$, respectively; for type 4, $\gamma = 1$ and $b = 1$. All streams consist of 50 jobs. The stream arrival rates for types 1, 2 and 3 are 0.1, 0.04 and 0.08 respectively, while the one for type 4 increases from 0.02 to about 0.2. The effect of these parameters is that the overall load on the system varies from about 60% to over 100%.

Fig. 5 shows the revenues achieved by the Current State heuristic, the Optimized Current State heuristic (as described in section 3.1), the Threshold heuristic (section 3.2) and the unrestricted 'Admit All' policy. Here, the charges and penalties differ from type to type: $c_1 = r_1 = 10$, $c_2 = r_2 = 20$, $c_3 = r_3 = 30$, $c_4 = r_4 = 40$.

The figure confirms that both heuristics cope well with increases in load (by rejecting more streams) and produce increasing revenues. Optimizing the server allocations achieves very small gains. In contrast, admitting all streams leads to a collapse of revenues.

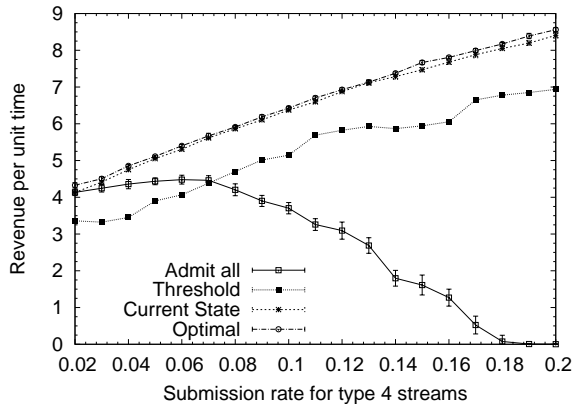

**Figure 5. Four service types; unequal charges.**

## 5. Conclusions

The contribution of this paper is to provide easily implementable policies for dynamically adaptable service provisioning systems subject to Quality of Service constraints and Service Level Agreements. We have demonstrated that decisions such as how many servers to allocate to each service type, and in particular whether to accept incoming streams or not, can have a significant effect on the revenue earned. Moreover, those decisions are affected by the contractual obligations between clients and provider in relation to quality of service.

On the basis of the limited experimentation performed so far, the Threshold heuristic would be a good candidate for practical implementation. The Current State heuristic has a slightly better performance, but is more demanding in terms of computational overheads. Of course, a more extensive programme of experimentation would be desirable before a definite conclusion is reached.

The following are some directions for future research.

1. It may be possible to share a server among several types of services. Then one would have to consider different job scheduling strategies, e.g., preemptive and non-preemptive priorities, Round-Robin, etc.

2. System reconfigurations, such as switching a server from one type of service to another, may incur non-negligible costs in either money or time. Taking those costs into account would mean dealing with a much more complex dynamic optimization problem.

The above avenues, and possibly others, are worth pursuing. In addition, if this methodology is to be applied in practice, it may be necessary to carry out some market research. It would be useful to discover what kind of waiting time obligations real users might ask for, and how much they would be willing to pay for them.

## Acknowledgements

This work was carried out as part of the research project QOSP (Quality Of Service Provisioning), funded by British Telecom. It was also supported by the European Union Network of Excellence EuroFGI (Future Generation Internet).